\documentclass{ws-procs975x65}
\usepackage{graphicx}
\usepackage{amsmath}

\def\beq{\begin{equation}}
\def\eeq{\end{equation}}

\begin{document}

\title{Gravity and the stability of the Higgs sector}

\author{Olga Czerwi{\'n}ska and Zygmunt Lalak}

\address{Institute of Theoretical Physics, Faculty of Physics, University of Warsaw,\\
ul. Pasteura 5, 02-093 Warsaw, Poland\\
E-mail: Olga.Czerwinska@fuw.edu.pl, Zygmunt.Lalak@fuw.edu.pl}

\author{{\L}ukasz Nakonieczny\footnote{Presenting author}}

\address{Institute of Theoretical Physics, Faculty of Physics, University of Warsaw,\\
ul. Pasteura 5, 02-093 Warsaw, Poland\\
E-mail: Lukasz.Nakonieczny@fuw.edu.pl}

\begin{abstract}
We pursued the question of the influence of a strong gravitational field on the structure of the
Higgs effective potential in the gauge-less top-Higgs sector of the Standard Model with an additional scalar singlet. 
To this end, we calculated the one-loop corrected effective potential in an
arbitrary curved spacetime. We have found that the gravity induced terms in the effective potential 
may influence its  behavior in both small and large field regions.
This result indicated the necessity of a~more careful investigation of the effect of high
curvature in the problems concerning the stability of the Higgs effective potential in the full Standard Model.
\end{abstract}


\bodymatter


\section{Introduction}

With the latest precise measurement of the top quark mass and properties of the Higgs boson, a~question of the 
stability of the Higgs effective potential received a~renewed attention. Using flat spacetime 
methods, various groups gauged the influence of the quantum effects on this subject, see for example   
\cite{Sher_1989,Buttazzo_Degrassi_Giardino_Giudice_Sala_Salvio_Strumia_2013,Lalak_Lewicki_Olszewski_2014} and the references therein.
On the other hand, since symptoms of the instability of the Standard Model (SM) vacuum appear at very high energy (of the order of $10^{10} \div 10^{11}$GeV), 
the effect of the gravity may be also important.

The aforementioned role of gravity in the problem of the stability was tackled, to some extent, using 
the effective operator approach \cite{Loebbert_Plefka_2015}. Unfortunately, the approach presented there, being based on a~non-covariant split of the spacetime metric 
on the Minkowski background and graviton fluctuations, may be unsuitable for the energy scale of the order of $10^{10}$GeV or above. 
One of the consequences of this may be the possibility that this method underestimates the importance of higher order curvature terms like
for example squares of the Ricci scalar, the Ricci tensor and the Riemann tensor which appear naturally if we demand a~one-loop renormalizability of the theory.
Recently, two other papers that consider the influence of gravity on the Higgs effective potential appeared \cite{Bezrukov_Rubio_Shaposhnikov_2014,Herranen_Markkanen_Nurmi_Rajantie_2015}. 
In the latter, only the tree-level potential was considered, while calculations in the former were based on the assumption of a~flat Minkowski background metric. For this reason, it was impossible to fully take into account the influence of the higher order curvature terms.    

After setting the problem of the stability of the SM vacuum in the frame of the hitherto existing attempts, let us discuss the  framework we used.
Although calculations in the full SM is an ultimate goal, technical difficulties caused by the presence of many interacting fields
are formidable. For this reason, we used a~simplified model consisting of a~real component of the Higgs field, top quark and an additional real scalar
singlet that may be considered as the mediator between the SM and the dark matter sector. We placed the matter model in the 
classical curved spacetime background that took the form of the Friedmann--Lema{\^i}tre--Robertson--Walker metric.
Using the heat kernel method (see for example \cite{DeWitt_1965,Buchbinder_Odintsov_Shapiro_1992}), we calculated the one-loop effective 
action for the considered model. This allowed us to find explicit forms of the gravity induced terms in the quantum corrected
effective potential and to assess its influence on the problem of the vacuum stability.

\section{Gravity and the stability of the scalar one-loop effective potential}

Let us start the discussion of our results by writing down the bare tree-level action of the theory that we considered
\begin{align}
\label{S_grav}
S_{grav} &= \int \sqrt{-g} d^4 x \bigg [
\frac{1}{16 \pi G_{B}} \left ( - R  - 2 \Lambda_B  \right ) + \nonumber \\
&+ \alpha_{1B} R_{\mu \nu \rho \sigma} R^{\mu \nu \rho \sigma} + \alpha_{2 B} R_{\mu \nu}R^{\mu \nu} + \alpha_{3B} R^2
\bigg ] ,\\
\label{S_scalar}
S_{scalar} &= \int \sqrt{-g} d^4 x \bigg [
  \nabla_{\mu} h_B  \nabla^{\mu} h_B -  m^2_{h B } h_B^2 + \xi_{h B }  h_B^2 R 
+  \nonumber \\  
&- \lambda_{h B}  h_B^4 - \lambda_{hXB} X_B^2 h_B^2     +  \nonumber \\  
&  +\nabla_{\mu} X_{B}  \nabla^{\mu} X_B -  m^2_{X B } X_B^2 + \xi_{X B } X_B^2 R
- \lambda_{X B}  X^4  
\bigg ], \\
\label{S_fermion}
S_{fermion} &=  \int \sqrt{-g} d^4 x \bigg [
\bar{ \psi}_{B} \left ( i \gamma^{\mu} \nabla_{\mu} - y_{B t} h_B \right ) \psi_{B} 
\bigg ],
\end{align}
where the subscript $B$ indicates bare quantities.\footnote{
We used the following sign conventions for the Minkowski metric tensor and the Riemann tensor:
\begin{align}
\eta_{a b} = diag(+,-,-,-), \quad
R_{\lambda \tau \mu }^{~~~~ \nu} = \partial_{\tau} \Gamma^{\nu}_{~~ \lambda \mu} + ... ~, \quad R_{\mu \nu} = R_{\mu \alpha \nu}^{~~~~ \alpha}. \nonumber 
\end{align}
}
In the above, $\psi$ is the Dirac type spinor representing the top quark field, namely $\psi = \left [ t_{L} , t_{R} \right ]^{T}$, $h$ is a~real scalar
representing the Higgs field, which corresponds to the radial mode of the full Higgs field in a~unitary gauge, and $X$ is an additional real scalar 
which was assumed to be the heavy mediator to the dark matter sector. 
Moreover, we also include terms that are of the second order in curvature in $S_{grav}$, they are needed for the renormalizability of
the theory (at the one-loop level), and also the nonminimal couplings of the scalar fields to gravity (terms $\xi_h h^2 R $ and $\xi_X X^2 R$) are needed for the same reason. 

Our main goal was to investigate the influence of the classical gravitational field on the one-loop effective potential.
To this end, we need to find both UV divergent terms (that give rise to the running of coupling constants as defined by the appropriate 
Renormalization Group Equations) and the finite part of the one-loop effective action. 
To accomplish both of these tasks we used the heat kernel method \cite{DeWitt_1965,Buchbinder_Odintsov_Shapiro_1992}, specifically the R-summed form of the 
proper time series representation of the heat kernel (the Schwinger-DeWitt series) \cite{Parker_Toms_1985}.
The details of the calculations were described in \cite{czerwinska_lalak_nakonieczny_2015}.
Below we present the approximation to the one-loop corrected effective potential to the scalar sector of the studied theory
\begin{align}
\label{one_loop_V}
- V^{(1)} &= - \frac{1}{2} \left ( m_h^2 - \xi_h R \right ) h^2 - \frac{\lambda_{h}}{4} h^4 - \frac{\lambda_{hX}}{4} h^2 X^2 
- \frac{1}{2} \left ( m_X^2 - \xi_X R \right ) X^2 - \frac{\lambda_{X}}{4} X^4 +   \nonumber \\
&+ \frac{\hbar}{64 \pi^2} \bigg [
 - tr \bigg ( a^2 \ln \Big ( \frac{a}{\mu^2} \Big ) \bigg ) + \frac{3}{2} tr a^2  + 8 b^2 \ln \Big ( \frac{b}{\mu^2} \Big ) - 12b^2 
+  \nonumber \\ 
&+ \frac{1}{3} y_t^2 h^2 \ln \Big ( \frac{b}{\mu^2} \Big ) R -  y_t^4 h^4 \ln \Big ( \frac{b}{\mu^2} \Big )  + \\
&- \frac{4}{180} \left ( - R_{\alpha \beta}R^{\alpha \beta} + R_{\alpha \beta \mu \nu} R^{\alpha \beta \mu \nu} \right ) \left ( \ln \Big ( \frac{a_{+}}{\mu^2} \Big ) 
+ \ln \Big (\frac{a_{-}}{\mu^2} \Big )  - 2 \ln \Big (\frac{b}{\mu^2} \Big ) \right )  + \nonumber \\
&- \frac{4}{3} R_{\alpha \beta \mu \nu} R^{\alpha \beta \mu \nu} \ln \Big (\frac{b}{\mu^2} \Big ) \nonumber 
\bigg ]
\bigg \},
\end{align}
where $a$ and $b$ are given by
\begin{align}
\label{b_entry}
b &= \frac{1}{2} y_t^2 h^2 - \frac{1}{12}R, \\
\label{a_entry}
a &= \begin{bmatrix}
 m_{X}^2 - ( \xi_{X}- \frac{1}{6} )R + 3 \lambda_{x} X^2 + \frac{ \lambda_{hX}}{2} h^2 & \lambda_{hX} h X \\
 \lambda_{hX} h X &  m_{h}^2 - ( \xi_{h} - \frac{1}{6} )R + 3 \lambda_{h} h^2 + \frac{ \lambda_{hX}}{2} X^2 \end{bmatrix}.
\end{align}
The eigenvalues of the matrix $a$ are denoted by $a_{\pm}$. Their exact forms can be straightforwardly obtained form (\ref{a_entry}).
All fields and coupling constants in the above expression are renormalized. As was mentioned, (\ref{one_loop_V}) represents an approximation to the 
effective potential which is valid when the spacetime curvature does not change very rapidly, $\nabla \nabla R << R^2$, and when it is not very big, $\frac{R}{m^2_{i}} << 1$,
where $m^2_{i} = \{b, a_{+}, a_{-}\}$ represents field dependent masses. It is straightforward to check that when we set all R-dependent terms in (\ref{one_loop_V})
to zero we obtain the standard Coleman-Weinberg form of the one-loop effective potential.  

To investigate the influence of the gravity induced terms we need to assume a~form of the spacetime metric. We choose the spatially flat Friedmann--Lema{\^i}tre--Robertson--Walker metric which gives us the following relation between energy density, pressure and spacetime curvature
\begin{align}
\label{R_rho}
&R = - 3 \bar{M_{P}}^{-2} \left [ -p + \frac{1}{3} \rho \right ], \nonumber \\
&- R_{\alpha \beta}R^{\alpha \beta } + R_{\alpha \beta \mu \nu} R^{\alpha \beta \mu \nu} = 2 \bar{M_{P}}^{-4} \rho \left ( \frac{1}{3} \rho + p \right ) , \\
&R_{\alpha \beta \mu \nu} R^{\alpha \beta \mu \nu} = 12 \bar{M_{P}}^{-4} \left [ \frac{1}{9} \rho^2 + \frac{1}{4} \left ( \frac{1}{3} \rho + p \right )^2 \right ], \nonumber 
\end{align}
where $\bar{M_{P}}$ is the reduced Planck mass.   

\begin{figure}[h]
\begin{center}
\includegraphics[width=0.7\textwidth]{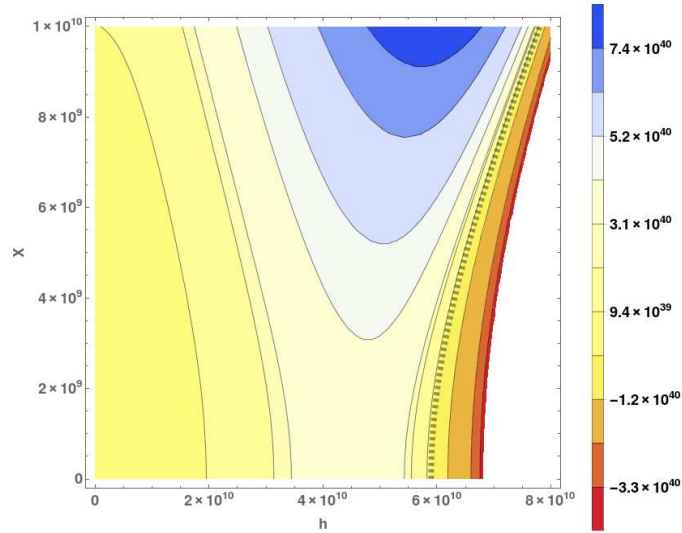}
\end{center}
\caption{Large field region of the effective one-loop potential of the scalar sector ($V^{(1)}$) in the radiation dominated era.
Total energy density was $\rho_{tot} = \sigma \nu^4 + \mu^4$, where $\mu = \frac{y_t}{\sqrt{2}} h$ is the running energy scale and we set $\sigma = 50$ and $\nu = 10^9$~GeV. 
The bold dashed line represents a~set of points at which $V^{(1)} = 0$ and to the right of this line the effective potential becomes negative.}
\label{fig1}
\end{figure}

In figure \ref{fig1} we present the shape of the one-loop effective potential in the scalar sector for the large field region.
We also take into account the one-loop running of the constants. From this figure we may see that the instability region, the border of which is marked by the bold dashed line, starts at the point $(X,h) \sim (0,10^{10})$ and then stretches to the larger value of the $X$ field. 

\section{Results}

In our investigations we ponder the question of the influence of the classical 
gravitational field on the form of the one-loop corrected scalar potential
for the gauge-less top-Higgs scalar mediator model. 
The approximation to the one-loop effective potential for the considered theory in the
curved spacetime background was obtained and presented above as (\ref{one_loop_V}). From it we may see that, within the range of the validity of our approximation,
the classical gravitational background will induce new terms in the aforementioned potential. In principle, these terms may influence both
the small field region (electroweak minimum) and also the large field one (the region for which the effective quartic coupling for the Higgs field becomes negative).

\begin{figure}[h]
\begin{center}
\includegraphics[width=0.7\textwidth]{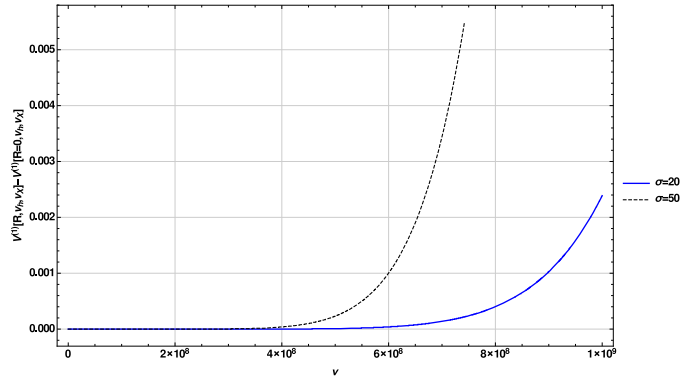}
\end{center}
\caption{Influence of the gravity induced terms on the electroweak minimum in the considered model. The running energy scale was set as $\mu = m_{top}$, where $m_{top}$ is the top quark mass and the total energy density
is given by the relation $\rho = \sigma \nu^4 + \mu^4$.}
\label{fig2}
\end{figure}

As far as the region of the electroweak minimum is concerned, the results for the radiation dominated era are presented
in figure \ref{fig2}. 
From the diagram we may infer that for the discussed setup gravity will contribute positively to the effective potential.
This means that electroweak minimum becomes shallower.   

On the other hand, to gauge the influence of gravity on the large field region we may do the following.
First of all, let us note that from figure \ref{fig1} we see that the instability region starts on the $X=0$ axis. 
Having this in mind, we may for now put $X=0$ in (\ref{one_loop_V}) and define the effective quartic coupling of the 
Higgs field in the presence of gravity as
\begin{align}
V^{(1)}(h) = \frac{1}{4} \left [ \lambda^{eff}_{h}(h)  + 
\frac{4}{64 \pi^2} \frac{1}{h^4} \frac{4}{3} R_{\alpha \beta \gamma \sigma} R^{\alpha \beta \gamma \sigma} \log{(\frac{y^2_{t} h^2}{2 \mu^2}})  \right ] h^4
\equiv \frac{1}{4} \bar{\lambda}^{eff}_{h}(R,h) h^4,
\end{align} 
where $\lambda^{eff}_{h}(h)$ is the effective field dependent quartic coupling defined in a~flat spacetime and $\bar{\lambda}^{eff}_{h}(R,h)$
is the field dependent coupling defined in the presence of gravity. In this definition we include only the term proportional 
to the Kretschmann scalar $\mathcal{K} =  R_{\alpha \beta \gamma \sigma} R^{\alpha \beta \gamma \sigma}$ since for the radiation dominated
era other terms are zero or negligible \cite{czerwinska_lalak_nakonieczny_2015}.  Since the new term contributes positively to the $\bar{\lambda}^{eff}_{h}(R,h)$,
we may ask the question how big energy density should be to slightly improve the behavior of $\lambda^{eff}_{h}(h)$?
Let us assume that for $h_{0} \sim 10^{10}$GeV the flat spacetime effective quartic coupling becomes small and negative. 
For concreteness, we took $\lambda^{eff}_{h}(h) = -0.02$. 
Using the relation (\ref{R_rho}) we may exchange $\mathcal{K}$ for the energy density (remembering that for the radiation dominated era we have $p = \frac{1}{3} \rho$)
and find the value of the critical~$\rho$ for which $\bar{\lambda}^{eff}_{h}(R,h) =0$. After an appropriate calculation one founds that 
\begin{align}
\rho_{crit} \approx \mu_{crit}^4 \Rightarrow \mu_{crit} \sim 10^{13} \div 10^{14} \textrm{GeV}.
\end{align}
This implies that if most of the energy of the matter system is stored in other field than the Higgs itself 
(which is a~typical situation in the radiation dominated era), then inclusion of the gravity induced terms in the effective potential 
may be of importance in analyzing the problem of the stability of the Standard Model.

\section*{Acknowledgments}
{\L}N was supported by the Polish National Science Centre under postdoctoral scholarship FUGA \mbox{DEC~--~2014/12/S/ST2/00332}.
ZL and OC were supported  by Polish National Science Centre under the research grant DEC-2012/04/A/ST2/00099.


\end{document}